\documentclass[11pt]{article}
\usepackage{amsfonts}
\usepackage{graphicx}
\usepackage{epstopdf}
\usepackage{epsfig}
\usepackage{amssymb}
\usepackage{setspace}
\usepackage{caption}
\usepackage{amsfonts}
\usepackage{color}
\usepackage{amsmath}
\usepackage{float}
\usepackage{comment}
\newcommand {\be}{\begin{equation}}
\newcommand {\ee}{\end{equation}}
 \newcommand {\bea}{\begin{array}}
 
 \newcommand {\eea}{\end{array}}

\textwidth=6.5in \hoffset=-.75in \voffset=-1in \numberwithin{equation}{section}
\numberwithin{figure}{section}

\begin{document}

\begin{titlepage}
	\vspace{1cm} 
	\begin{center}
		{\Large \bf {Accelerating black holes in the low energy heterotic string theory}}\\
	\end{center}
	\vspace{2cm}
	\begin{center}
		\renewcommand{\thefootnote}{\fnsymbol{footnote}}
		Haryanto M. Siahaan{\footnote{haryanto.siahaan@gmail.com}}\\~~\\
		MTA Lend\"{u}let Holographic QFT Group, Wigner Research Centre for Physics,\\
		Konkoly-Thege Mikl\'{o}s u. 29-33, 1121 Budapest, Hungary\\and\\
		Center for Theoretical Physics, Physics Department, Parahyangan Catholic University,\\
		Jalan Ciumbuleuit 94, Bandung 40141, Indonesia
		\renewcommand{\thefootnote}{\arabic{footnote}}
	\end{center}
	
	\begin{abstract}
We present a new solution describing a pair of rotating, charged, and accelerating black holes in in the low energy limit of heterotic string theory. To obtain the solution, we apply the Hassan-Sen transformation to the neutral accelerating Kerr spacetime as the seed solution. Some properties of the black hole solutions are discussed.
	\end{abstract}
\end{titlepage}\onecolumn 
\bigskip 

\section{Introduction}
\label{sec:intro}

String theory is considered by many as a consistent explanation of gravity in quantum level. That is why, the discussions of some low energy limits of string theory which contain gravity are still among the main present interests in gravitational physics studies. In the heterotic string theory, the low energy limit has graviton, Abelian gauge field, dilaton, and second rank antisymmetric tensor fields, and the corresponding rotating and charged black hole solution was given by Sen \cite{Sen:1992ua}, namely Kerr-Sen black hole. This Kerr-Sen black hole has mass, rotation, and electric charge, which make it is analogous to the well-known Kerr-Newman black hole in Einstein-Maxwell theory. Despite their similarities, Kerr-Sen and Kerr-Newman black holes can be distinguished in some particular ways, for example the strong light deflection limit in Kerr-Sen background \cite{Gyulchev:2006zg} and the hidden conformal symmetry \cite{Ghezelbash:2012qn}. These distinctions between Kerr-Sen and Kerr-Newman black holes, in addition to the expectation that string theory is the ultimate explanation of all fundamental processes, motivate researches to explore more on the properties of Kerr-Sen black holes \cite{Siahaan:2015ljs,Siahaan:2015xna,Bernard:2017rcw,Bernard:2016wqo,Huang:2017whw,Uniyal:2017yll,Liu:2018vea}.

In obtaining Kerr-Sen solution, Sen employed the Hassan-Sen transformation \cite{Hassan:1991mq} which based on the symmetry in the low energy limit of heterotic string theory action. This Hassan-Sen transformation maps a known solution in the theory, namely the seed solution, to another one which belongs to the same theory. Nevertheless, normally one would consider a quite simple seed solution, since the mapping can lead to a complicated expression for all incorporated fields. In Kerr-Sen case \cite{Sen:1992ua}, Sen employed the Kerr metric as the seed solution to obtain a set of non-trivial solutions for all fields in the theory. Shortly after, the solution generating method was used in \cite{Johnson:1994ek} where the authors considered the non-rotating Taub-NUT spacetime as the seed solution. 

In the beginning of black hole studies, these objects were considered merely as mathematical entity rather than real existing part of the universe. Nowadays, especially after the successful detection of gravitational waves from black holes collisions \cite{TheLIGOScientific:2016src} and the possibility of looking at an astrophysical black holes \cite{Ricarte:2014nca}, black holes have become a central in the current physics research. As a theoretical object, black hole is used an arena to reconcile gravity and quantum mechanics. Therefore, understanding a black hole solution is important, from which the behavior of the black hole can be figured out. Nevertheless, not all exact black holes solutions are fully understood, even in the vacuum Einstein theory. For example the $C$-metric, which is interpreted as spacetime solution describing two masses causally separated and constantly accelerating away each other. The mass and thermodynamics in this spacetime are still under investigations \cite{Appels:2016uha,Astorino:2016ybm}.

The rotating $C$-metric, or sometime referred as the accelerating Kerr spacetime, also solves the vacuum Einstein equations, even when the spacetime equipped with the NUT charge. Thence, one might wonder what happens if the Hassan-Sen transformation is applied to this accelerating Kerr solution. One could expect to get a set of fields solutions describing the accelerating version of Kerr-Sen black hole, analogous to the accelerating Kerr-Newman in the Einstein-Maxwell theory \cite{Griffiths:2009dfa}. This is exactly what we will perform in this paper, applying the Hassan-Sen transformation to the accelerating Kerr(-NUT) spacetime to get the solutions to fields that obey the equations of motion in the low energy limit of heterotic string theory. 

The organization of this paper is as follows. In the next section, we provide a review on the Hassan-Sen transformation with explicit results to a stationary and axially symmetric seed solution of the vacuum Einstein equations. In this section we also survey briefly the accelerating black holes in the vacuum Einstein system. In section \ref{s.1}, we present the accelerating black hole solutions in heterotic string theory, presented in both Boyer-Lindquist-like and Plebanski-Demianski coordinates. In the next section, some physical properties of the obtained black hole solutions are discussed. Finally, discussions and conclusions are given in the next section. We also provide appendices discussing the Hassan-Sen transformation which include the NUT charge, and the charges formulas in the low energy heterotic string theory.

\section{Preliminaries on Hassan-Sen transformation and accelerating black holes}

\subsection{Hassan-Sen transformation in the low energy limit of heterotic string field theory}\label{revHS}

In \cite{Sen:1992ua}, Sen transformed the Kerr solution to get a charged and rotating black hole spacetime in the low energy limit of heterotic string theory. The transformation employed is the one introduced by Hassan and Sen\footnote{In this paper would be referred as the Hassan-Sen transformation.} \cite{Hassan:1991mq} that transforms a set of solution $\left\{ {{\tilde g}_{\mu \nu } ,{\tilde A}_\mu  ,{\tilde \Phi} ,{\tilde B}_{\mu \nu } } \right\}$ to another one $\left\{ {g_{\mu \nu }, A_\mu  ,\Phi, B_{\mu \nu } } \right\}$ which solve the equations of motion from the action
\be\label{action.het}
S = \int {d^4 x} \sqrt { - g} e^{ - \Phi } \left( {R + \left( {\nabla \Phi } \right)^2  - \frac{1}{8}F_{\mu \nu } F^{\mu \nu }  - \frac{1}{{12}}H_{\alpha \beta \gamma } H^{\alpha \beta \gamma } } \right)\,.
\ee
In the equation above, $H_{\alpha \beta \gamma }  = \partial _\alpha  B_{\beta \gamma }  + \partial _\gamma  B_{\alpha \beta }  + \partial _\beta  B_{\gamma \alpha }  - \frac{1}{4}\left( {A_\alpha  F_{\beta \gamma }  + A_\gamma  F_{\alpha \beta }  + A_\beta  F_{\gamma \alpha } } \right)$ and $F_{\mu \nu }  = \partial _\mu  A_\nu   - \partial _\nu  A_\mu  $. When all non-gravitational fields $\left\{ {A_\mu  ,\Phi ,B_{\mu \nu } } \right\}$ vanish, then the system described by action (\ref{action.het}) is just the vacuum Einstein. Thats why, the Kerr spacetime which solves the vacuum Einstein equations can be transformed using Hassan-Sen transformation to get a new set of fields that belongs to the action (\ref{action.het}), where some components of the non-gravitational fields are non-zero.

Here let us review the Hassan-Sen transformation applied explicitly to a stationary and axial symmetric solution to the vacuum Einstein equations, specifically in the metric tensor expression
\be\label{metric0}
ds^2  = g_{tt} dt^2  + 2g_{t\phi } dtd\phi  + g_{\phi \phi } d\phi ^2  + g_{rr} dr^2  + g_{xx} dx^2 \,.
\ee 
Here we are using the coordinate $\left(t,r,x,\phi\right)$, where the stationary and axial symmetry require $g_{\mu\nu} = g_{\mu\nu} \left(r,x\right)$. In this review, we aim to map $\left\{ {{\tilde g}_{\mu \nu } } \right\}$ with the action
\be\label{action.vacuum.Einstein}
S = \int {d^4 x\sqrt { - {\tilde g}} {\tilde R}} 
\ee 
to the solutions $\left\{ {g_{\mu \nu } ,A_\mu  ,\Phi ,B_{\mu \nu } } \right\}$ of the action
\be\label{action.het.capital}
S = \int {d^4 x} \sqrt { - g} e^{ - \Phi } \left( {R+ g^{\mu \nu } \partial _\mu  \Phi \partial _\nu  \Phi  - \frac{1}{8}g^{\mu \alpha } g^{\nu \beta } F_{\mu \nu } F_{\alpha \beta }  - \frac{1}{{12}}g^{\mu \alpha } g^{\nu \beta } g^{\gamma \lambda } H_{\mu \nu \gamma } H_{\alpha \beta \lambda } } \right)\,.
\ee 
From now on, we denote the fields with ``tilde'' as the seed solutions, and the ones with no ``tilde'' as the Hassan-Sen transformed result. Equations of motion derived from the action (\ref{action.het.capital}) are
\be \label{eqG}
E_{\mu \nu }  + \nabla _\mu  \nabla _\nu  \Phi  + \frac{1}{2}g_{\mu \nu } \left( {\left( {\nabla \Phi } \right)^2  - 2\nabla ^2 \Phi } \right) = \frac{1}{4}\left[ {F_{\mu \alpha } F_\nu ^\alpha   + H_{\mu \alpha \beta } H_\nu ^{\alpha \beta }  - \frac{1}{2}g_{\mu \nu } \left( {\frac{{F^2 }}{2} + \frac{{H^2 }}{3}} \right)} \right]\,,
\ee 
\be \label{eqPh}
\left( {\nabla \Phi } \right)^2  - 2\nabla ^2 \Phi  = R - \frac{{F^2 }}{8} - \frac{{H^2 }}{{12}}\,,
\ee
\be \label{eqA}
\nabla _\mu  F^{\mu \nu }  = F^{\mu \nu } \partial _\mu  \Phi  + \frac{1}{2}F_{\alpha \beta } H^{\nu \alpha \beta } \,,
\ee
and
\be \label{eqB}
\nabla _\alpha  H^{\alpha \mu \nu }  = H^{\alpha \mu \nu } \partial _\alpha  \Phi \,.
\ee
In the equations above, $E_{\mu\nu}$ is the Einstein tensor with spacetime metric $g_{\mu\nu}$. Interestingly, the solutions to these equations of motion can describe black holes, such as Kerr-Sen \cite{Sen:1992ua} and Gibbons-Maeda-Garfinkle-Horowitz-Strominger \cite{Gibbons:1987ps,Garfinkle:1990qj} black holes. 
 
To proceed the Hassan-Sen transformation, we first introduce a second rank tensor $K_{\mu\nu}$ defined by
\be
K_{\mu \nu }  =  - g_{\mu \nu }  - \frac{1}{4}A_\mu  A_\nu   - B_{\mu \nu }
\ee 
whose matrix expression is $\bf{K}$. Then let us denote the flat spacetime metric 
\be 
h_{\mu\nu} = {\rm diag}\left(1,1,1,-1\right)
\ee 
which is $\bf h$ in the matrix form. Now we can construct a $9\times 9$ matrix
\[
{\bf M} = \left( {\begin{array}{*{20}c}
	{\left( {{\bf K}^T  - {\bf h} } \right){\bf g}^{-1} \left( {{\bf K} - {\bf h} } \right)} & {\left( {{\bf K}^T  - {\bf h} } \right){\bf g}^{-1} \left( {{\bf K} + {\bf h} } \right)} & { - \left( {{\bf K}^T  - {\bf h} } \right){\bf g}^{-1} {\bf A}}  \\
	{\left( {{\bf K}^T  + {\bf h} } \right){\bf g}^{-1} \left( {{\bf K} - {\bf h} } \right)} & {\left( {{\bf K}^T  + {\bf h} } \right){\bf g}^{-1} \left( {{\bf K} + {\bf h} } \right)} & { - \left( {{\bf K}^T  + {\bf h} } \right){\bf g}^{-1} {\bf A}}  \\
	{ - {\bf A}^T {\bf g}^{-1} \left( {{\bf K} - {\bf h} } \right)} & { - {\bf A}^T {\bf g}^{-1} \left( {{\bf K} + {\bf h} } \right)} & {{\bf A}^T {\bf g}^{-1} {\bf A}}  \\
	\end{array}} \right)
\]
which contains all the fields in the theory except the dilaton $\Phi$. Intuitively, the matrix $\bf A$ above is the column vector expression of $A_\mu$. It is found that the action (\ref{action.het}) is invariant under the transformation \cite{Hassan:1991mq}
\be 
{\bf {\tilde M}} \to {\bf M} = {\bf \Theta {\tilde M}\Theta} ^T \,,
\ee
together with the change of dilaton field
\be 
{\tilde \Phi}  \to \Phi  = \Phi  + \frac{1}{2}\ln \frac{{\det {\bf g}}}{{\det {\bf {\tilde g}}}}\,,
\ee
where
\be 
{\bf \Theta}  = \left( {\begin{array}{*{20}c}
	{I_{7 \times 7} } & {...} & {...}  \\
	{...} & {\sqrt {1 + s^2 } } & s  \\
	{...} & s & {\sqrt {1 + s^2 } }  \\
	\end{array}} \right)\,.
\ee
Above, it is intuitive that all matrices with ``tilde'' contains the seed fields solutions, and the matrices are represented by the boldface letters. The dots in $\bf \Theta$ stand for zero component in the matrix, and $I_{7 \times 7}$ stands for the seven dimensional identity matrix, and $s$ is some arbitrary real parameter.

Since we start with all non-gravitational fields vanish, then the Hassan-Sen transformed field solutions would depend on the metric tensor ${\tilde g}_{\mu\nu}$ only. The results are can be found as follows. The metric components $g_{\mu\nu}$ expressed in tetrads are given by
\[
e_\mu ^{\left( 0 \right)} dx^\mu   = \frac{{\sqrt { - {\tilde g}_{tt} } }}{\Lambda }\left( {dt + \frac{{{\tilde g}_{t\phi } }}{{{\tilde g}_{tt} }}\left( {1 + s^2 } \right)d\phi } \right)\,,
\]
\be\label{tetrad.HassanSen}
e_\mu ^{\left( 1 \right)} dx^\mu   = \sqrt {{\tilde g}_{rr} } dr
~~,~~
e_\mu ^{\left( 2 \right)} dx^\mu   = \sqrt {{\tilde g}_{xx} } dx
~~,~~
e_\mu ^{\left( 3 \right)} dx^\mu   = \sqrt {{\tilde g}_{\phi \phi }  - \frac{{{\tilde g}_{t\phi }^2 }}{{{\tilde g}_{tt} }}} d\phi\,,
\ee 
where $g_{\mu \nu }  = \eta _{\left( a \right)\left( b \right)} e_\mu ^{\left( a \right)} e_\nu ^{\left( b \right)} $ and $\Lambda = 1+s^2\left(1+{\tilde g}_{tt}\right)$. Here we use the flat metric
\be 
\eta _{\left( a \right)\left( b \right)}  = \left( {\begin{array}{*{20}c}
		{ - 1} & 0 & 0 & 0  \\
		0 & 1 & 0 & 0  \\
		0 & 0 & 1 & 0  \\
		0 & 0 & 0 & 1  \\
\end{array}} \right)\,.
\ee
The gauge field and dilaton fields are given by 
\be \label{vector.HassanSen}
A_\mu  dx^\mu   = \frac{{2s\sqrt {1 + s^2 } }}{\Lambda }\left( {\left( {1 + {\tilde g}_{tt} } \right)dt + {\tilde g}_{t\phi } d\phi } \right)\,,
\ee
\be \label{dilaton.HassanSen}
\Phi  =  - \ln \Lambda \,,
\ee 
respectively. Finally, the only non-vanishing second rank tensor field's component is found to be
\be \label{secondrank.HassanSen}
B_{t\phi }  =  - B_{\phi t}  = \frac{{s^2 {\tilde g}_{t\phi } }}{\Lambda }\,.
\ee
So now we have a general formulas of all fields $\left\{ {g_{\mu \nu } ,A_\mu  ,\Phi ,B_{\mu \nu } } \right\}$ in the action (\ref{action.het.capital}) as some functions of a vacuum Einstein solution (\ref{metric0}). When the parameter $s$ vanishes, all the non-gravitational fields (\ref{vector.HassanSen}) - (\ref{secondrank.HassanSen}) are zero, since the transformation matrix ${\bf \Theta}$ would be just an identity.

\subsection{Accelerating Kerr black holes}

In a vacuum Einstein system, there exist spacetime solution which is interpreted to contain a pair of black holes accelerating away each other at a constant rate. This solution is known as the $C$-metric \cite{Kinnersley:1970zw,Weyl:1917gp}, which can be considered as a special case of a more general solution in Einstein-Maxwell theory known by Plebanski-Demianski (PD) solution \cite{Plebanski:1976gy}. Restricted in the vacuum case, the most general expression for accelerating spacetime is given in \cite{Griffiths:2006tk}, where they perform rescaling to the original PD metric and introduce two additional real parameters, $\alpha$ and $\omega$. The line element is
\be \label{metricGP}
ds^2  = \frac{1}{{\left( {1 + \alpha p r} \right)^2 }}\left[ { - \frac{Q}{{\varrho ^2 }}\left( {d\tau  - \omega p^2 d\varphi } \right)^2  + \varrho ^2 \left( {\frac{{dr^2 }}{Q} + \frac{{dp^2 }}{K}} \right) + \frac{K}{{\varrho ^2 }}\left( {\omega d\tau  + r^2 d\varphi } \right)^2 } \right]
\ee
where
\be \label{PnoNUT}
K = k + 2np \omega ^{ - 1}  - \varepsilon p^2  - 2\alpha m p^3  - \alpha ^2 \omega ^2 k p^4 \,,
\ee 
\be\label{QnoNUT} 
Q = \omega ^2 k - 2mr + \varepsilon r^2  + 2\alpha nr^3 \omega ^{ - 1}  - \alpha ^2 kr^4 \,,
\ee
and $\varrho^2 = r^2 + \omega^2 p^2$. In the expressions above, $m$, $n$, $\epsilon$, $k$, $\alpha$, and $\omega$ are some real parameters. This metric solves the vacuum Einstein equation, and in some literature it is argued that $n$ is related the NUT parameter \cite{Hong:2004dm}. However, Griffiths and Podolsky showed that to get the spacetime equipped with a NUT parameter, a further set of coordinates transformation applied to (\ref{metricGP}) is needed. In this coordinate transformation, a new parameter $l$ is introduced, and the transformation between $\left\{ {\tau ,r,p,\varphi } \right\}$ and $\left\{ {t,r,x,\phi } \right\}$ reads
\be \label{NUTtransform}
p \to \frac{{\left( {l + ax} \right)}}{\omega }~~,~~\tau  \to t - \frac{{\left( {l + a} \right)^2 }}{a}\phi ~~,~~\varphi  \to  - \frac{\omega }{a}\phi \,.
\ee
Applying (\ref{NUTtransform}) to the metric (\ref{metricGP}) yields
\[ 
ds^2  = \frac{1}{{ \Omega ^2 }}\left[ { - \frac{{ Q}}{{ \varrho ^2 }}\left( {dt - \left( {a\left( {1 - x^2 } \right) + 2l\left( {1 - x} \right)} \right)d\phi } \right)^2  +  \varrho ^2 \left( {\frac{{dr^2 }}{{Q}} + \frac{{a^2 dx^2 }}{{\omega^2 P}}} \right)} \right.
\]
\be \label{metricrotCNUT}
\left. { + \frac{{ P}}{{ \varrho ^2 }}\left( {\omega dt - \frac{{\omega \left( {r^2  + \left( {l + a} \right)^2 } \right)}}{a}d\phi  } \right)^2 } \right]
\ee
where
\be 
\Omega  = 1 + \alpha \left( {l + ax} \right)r\omega ^{ - 1} \,,
\ee 
\be 
\varrho ^2  = r^2  + \left( {l + ax} \right)^2 \,,
\ee
and  
\be 
P = c_0  + c_1 x + c_2 x^2  + c_3 x^3  + c_4 x^4 \,,
\ee 
where the coefficients $c_i$ can be written as
\[
c_0  = k - l\omega ^{ - 2} \left( {\alpha ^2 kl^3  + \varepsilon l - 2n} \right) - 2\alpha ml^3 \omega ^{ - 3} \,,
\]
\[
c_1  = 2a\omega ^{ - 2} \left( {n - \varepsilon l - 2\alpha ^2 kl^3 } \right) - 6\alpha ml^2 a\omega ^{ - 3} \,,
\]
\[
c_2  =  - a^2 \omega ^{ - 2} \left( {\varepsilon  + 6\alpha ^2 kl^2 } \right) - 6\alpha mla^2 \omega ^{ - 3} \,,
\]
\[
c_3  =  - 2\alpha a^3 \omega ^{ - 3} \left( {m + 2\alpha kl\omega } \right)\,,
\]
\[
c_4  =  - \alpha ^2 ka^4 \omega ^{ - 2} \,.
\]
The Kerr-NUT spacetime can be extracted fromt (\ref{metricrotCNUT}) by setting $\alpha = 0$, $k = \left(a^2-l^2\right)\omega^{-2}$, $\epsilon = 1$, $n=l$, and $\omega = a$,
\[
ds^2  =  - \frac{{\left( {r^2  - 2mr + a^2  - l^2 } \right)}}{{r^2  + \left( {l + ax} \right)^2 }}\left\{ {dt - \left[ {a\left( {1 - x^2 } \right) + 2l\left( {1 - x} \right)} \right]d\phi } \right\}^2  + \frac{{\left( {r^2  + \left( {l + ax} \right)^2 } \right)dx^2 }}{{1 - x^2 }}
\]
\be \label{metricKerrNUT}
+ \frac{{\left( {r^2  + \left( {l + ax} \right)^2 } \right)dr^2 }}{{\left( {r^2  - 2mr + a^2  - l^2 } \right)}} + \frac{{\left( {1 - x^2 } \right)}}{{r^2  + \left( {l + ax} \right)^2 }}\left\{ {adt - \left[ {r^2  + \left( {a + l} \right)^2 } \right]d\phi } \right\}^2 \,.
\ee 
At this point we can see how the new parameter $n$ is related to the NUT charge $l$ in the Kerr-NUT metric (\ref{metricKerrNUT}), from which we can learn that $n$ is not simply the NUT charge, but related to it\footnote{In the other words, one can say that setting the acceleration parameter $\alpha$ to be zero in (\ref{metricGP}) does not lead to the proper Kerr-NUT line element (\ref{metricKerrNUT}) directly.} \cite{Griffiths:2005se}. Note that all the metrics (\ref{metricGP}), (\ref{metricrotCNUT}), and (\ref{metricKerrNUT}) fall into the category of spacetime (\ref{metric0}), and also solve the vacuum Einstein equations. Therefore, the Hassan-Sen transformation discussed previously apply to all of these metrics, and the resulting set of solutions would be a new fields in the low energy limit of heterotic string theory. Nevertheless, the case that we have particular interest and becomes the main discussion in this paper is the special limit of (\ref{metricrotCNUT}) describing accelerating Kerr black holes without NUT charge, i.e. $l=0$. This is what we will elaborate in the next section. 

\section{Hassan-Sen transformation to accelerating black holes}\label{s.1}

\subsection{Spherical-type coordinate}

The rotating and accelerating black hole solution in vacuum Einstein theory can be obtained from the general line element (\ref{metricrotCNUT}) by setting 
\be \label{trans.1.PDtoHT}
l=0~~,~~ k=1~~,~~ \omega = a~~,~~ n = m\alpha a~~,~~ \epsilon = -\alpha^2 a^2\,,
\ee 
followed by the ``shifts''
\be \label{trans.2.PDtoHT}
P \to {\cal P} = P - x^2  ~~,~~ Q \to {\cal Q} = Q + r^2 \,.
\ee 
After performing this transformation, we have
\be \label{rotcmetric}
ds^2  = \Omega ^{ - 2} \left[ { - \frac{\cal Q}{{\rho ^2 }}\left( {dt - a \left(1-x^2\right) d\phi } \right)^2  + \frac{{\rho ^2 dr^2 }}{\cal Q} + \frac{{\rho ^2 dx^2 }}{{{\cal P}}} + \frac{{{\cal P}}}{{\rho ^2 }}\left( {adt - \left(r^2+a^2\right) d\phi } \right)^2 } \right]
\ee 
where
$\Omega  = 1 + \alpha rx$, $\rho^2  = r^2  + a^2 x^2 $,
\be {\cal P} = \left(1 + 2\alpha mx + \alpha ^2 a^2 x^2\right)\left(1-x^2\right)\,, \ee 
and 
\be 
{\cal Q} = \left( {r - r_ +  } \right)\left( {r - r_ -  } \right)\left( {1 - \alpha ^2 r^2 } \right)\,.
\ee  
In the equation above, $r_\pm$ is the roots of $r^2  - 2mr + a^2  = 0$, i.e. the outer and inner horizons of Kerr black hole, and we consider $m > a$. This spacetime solution is interpreted to describe a pair of rotating neutral black holes which are constantly accelerating each other with the acceleration magnitude $a_\mu  a^\mu   =  - \alpha ^2 $ \cite{Griffiths:2009dfa}. Setting $\alpha = 0$ in (\ref{rotcmetric}) yields the Kerr solution, and the case of $a=0$ in this line element is the $C$-metric.

Since the metric (\ref{rotcmetric}) has the form of (\ref{metric0}), then one can directly show the corresponding line element solution (\ref{tetrad.HassanSen}) as a result of Hassan-Sen transformation to be
\[
ds_{string}^2  = g_{\mu \nu } dx^\mu  dx^\nu  =  - \frac{{{\cal Q} - {\cal P}a^2 }}{{\Omega ^2 \rho ^2 \Lambda ^2 }}\left( {dt - \frac{{a\left( {1 + s^2 } \right)\left( {{\cal Q}{\left(1-x^2\right)}  - \left(r^2+a^2\right) {\cal P}} \right)}}{{{\cal Q} - {\cal P}a^2 }}d\phi } \right)^2 
\]
\be \label{CKSstringframe}
+ \frac{{\rho ^2 }}{{\Omega ^2 }}\left( {\frac{{dr^2 }}{\cal Q} + \frac{{dx^2 }}{{\cal P}} + \frac{{{\cal QP}}}{{{\cal Q} - {\cal P}a^2 }}d\phi ^2 } \right)\,.
\ee
Accordingly, the vector (\ref{vector.HassanSen}), dilaton (\ref{dilaton.HassanSen}), and second rank tensor fields (\ref{secondrank.HassanSen}) read
\be \label{solAht}
A_\mu  dx^\mu   = \frac{{2s\sqrt {1 + s^2 } }}{\Omega ^2 \rho^2\Lambda}\left( {{\left({\Omega ^2 \rho ^2  - {\cal Q} + {\cal P}a^2 }\right)}dt + {a\left( {{\cal Q}\left(1-x^2\right)  - \left(r^2+a^2\right) {\cal P}} \right)}d\phi } \right) \,,
\ee
\be\Phi  =  - \ln \Lambda \,,
\ee 
and
\be \label{solBht}
B_{t\phi }  =  - B_{\phi t}  = \frac{{s^2 a\left( {{\cal Q}\left(1-x^2\right)  - \left(r^2+a^2\right) {\cal P}} \right)}}{{\Lambda \Omega ^2 \rho ^2 }}\,,
\ee
respectively. In the expressions above we have used
\be 
\Lambda  = 1 + s^2 \left( {1 - \frac{{{\cal Q} - {\cal P}a^2 }}{{\Omega ^2 \rho ^2 }}} \right)\,.
\ee

Alternatively, the set of solutions (\ref{CKSstringframe})-(\ref{solBht}) can also be obtained by performing the Hassan-Sen transformation to the metric (\ref{metricrotCNUT}), and employ the conditions (\ref{trans.1.PDtoHT}) and (\ref{trans.2.PDtoHT}) to the obtained set of solutions. The set of solutions (\ref{CKSEinsteinframe}) - (\ref{solBht}) reduce to that of Kerr-Sen system \cite{Sen:1992ua} for $\alpha =0$, and setting $s=0$ to (\ref{CKSEinsteinframe}) - (\ref{solBht}) yields the vanishing of all non-gravitational fields representing the generic accelerating Kerr spacetime in vacuum Einstein. Furthermore, for the discussions on some physical aspects of the black hole, let us express the metric (\ref{CKSEinsteinframe}) in the Einstein frame given by
\be \label{stringTOeinsteinFRAME}
ds_E^2  = \exp \left( { - \Phi } \right)ds_{string}^2\,,
\ee 
which reads
\[
ds_{E}^2  =  - \frac{{{\cal Q} - {\cal P}a^2 }}{{\Omega ^2 \rho ^2 \Lambda }}\left( {dt - \frac{{a\left( {1 + s^2 } \right)\left( {{\cal Q}{\left(1-x^2\right)}  - \left(r^2+a^2\right) {\cal P}} \right)}}{{{\cal Q} - {\cal P}a^2 }}d\phi } \right)^2 
\]
\be \label{CKSEinsteinframe}
+ \frac{{\rho ^2 \Lambda}}{{\Omega ^2 }}\left( {\frac{{dr^2 }}{\cal Q} + \frac{{dx^2 }}{{\cal P}} + \frac{{{\cal QP}}}{{{\cal Q} - {\cal P}a^2 }}d\phi ^2 } \right)\,.
\ee

In the next section we will present the result of Hassan-Sen transformation to the accelerating Kerr solution expressed in the Plebanski-Demianski form. In the present form (\ref{CKSEinsteinframe}), an advantage is the Boyer-Lindquist-like expression which normally preferred when one discusses some practical aspects of the black hole, such as the kinematics of a test particle \cite{Siahaan:2015ljs} and light deflection \cite{Gyulchev:2006zg}. However, some properties of the solutions such as the regularity and spacetime signature will become more handy if the line element is expressed in the Plebanski-Demianski form. This is what we will do next.

\subsection{Plebanski-Demianski-type coordinate}

Another popular expression for accelerating black hole spacetime is in the form of Plebanski and Demianski \cite{Plebanski:1976gy}, or the one similar to that\footnote{Related to the original Plebanski and Demianski form after some scalings.}, which can be obtained from (\ref{rotcmetric}) by performing the transformations $r\to -\alpha^{-1}y^{-1}$ and $t \to \alpha^{-1} t$. The result reads
\be\label{CmetricXYkerr}
ds^2  = \frac{1}{{\Xi ^2 }}\left[{\frac{G_{\left(y\right)}}{\cal F} {\left( {dt - a\alpha \left(1-x^2\right) d\phi } \right)^2  + \frac{G_{\left(x\right)}}{\cal F} \left( { \left(1+a^2\alpha^2 y^2\right) d\phi  - a\alpha y^2 dt} \right)^2 }} {+ {\cal F }\left( {\frac{{dx^2 }}{{G_{\left( x \right)}}} - \frac{{dy^2 }}{{G_{\left( y \right)}}}} \right)}\right]
\ee 
where
\be 
\Xi = \alpha \left( {x - y} \right)
~~,~~
{\cal F }= 1 + \left( {a\alpha xy} \right)^2\,,
\ee 
and the structure function \cite{Hong:2003gx,Hong:2004dm}
\be\label{structureF} 
G_{\left( \xi  \right)}  = \left( {1 - \xi ^2 } \right)\left( {1 + \xi \alpha r_ -  } \right)\left( {1 + \xi \alpha r_ +  } \right)\,.
\ee
Here, $r_\pm$ are the same to those in (\ref{rotcmetric}). However, the non-accelerating version of (\ref{CmetricXYkerr}) cannot be simply obtained by taking $\alpha \to 0$, unlike in the case (\ref{rotcmetric}).

Spacetime expressed in the form of (\ref{CmetricXYkerr}) is claimed to be free of no closed timelike curve (CTC) and torsion singularity \cite{Hong:2004dm,Bonnor:2002fk}. The factorizable of structure functions (\ref{structureF}) is due to the re-expression by Hong and Teo \cite{Hong:2003gx,Hong:2004dm}. In the form of (\ref{structureF}), it is easier to analyze the spacetime properties, for example the condition to maintain the correct spacetime signature \cite{Griffiths:2009dfa}. This section is devoted to present the solutions (\ref{tetrad.HassanSen}) - (\ref{secondrank.HassanSen}) in terms of seed metric (\ref{CmetricXYkerr}). The set of solutions $\left\{ {g_{\mu \nu } ,\Phi ,A_\mu  ,B_{\mu \nu } } \right\}$ can be written as the followings.
\[
ds_{string}^2  = \frac{{{\cal F}\Xi ^2 \Upsilon }}{{{\cal F}\Xi ^2 \left( {1 + s^2 } \right) + s^2 \Upsilon }}\left( {dt + \frac{{a\alpha \left( {1 + s^2 } \right)\left( {\Upsilon + G_{\left( x \right)} y^2  - G_{\left( y \right)} x^2 } \right)d\phi}}{\Upsilon }} \right)^2 
\]
\be\label{CmetricXYhet}
+ \frac{{\cal F}}{{\Xi ^2 }}\left( {\frac{{dx^2 }}{{G_{\left( x \right)} }} - \frac{{dy^2 }}{{G_{\left( y \right)} }} + \frac{{G_{\left( x \right)} G_{\left( y \right)} d\phi ^2 }}{\Upsilon }} \right)\,,
\ee
\be \label{PhiXYhet}
\Phi  = - \ln \left( \frac{{\cal F}\Xi ^2  + s^2 \left( {{\cal F}\Xi ^2  + \Upsilon } \right)}{{\cal F}\Xi ^2 } \right)\,,
\ee

\be\label{AvecXYhet}
A_\mu  dx^\mu   = \frac{{2s\sqrt {1 + s^2 } \left( {\left( {{\cal F}\Xi ^2  + \Upsilon } \right)dt - a\alpha \left( {\Upsilon + G_{\left( x \right)} y^2  - G_{\left( y \right)} x^2 } \right)d\phi } \right)}}{{{\cal F}\Xi ^2  + s^2 \left( {{\cal F}\Xi ^2  + \Upsilon } \right)}}\,,
\ee
and
\be \label{BfXYhet}
B_{t\phi }  =  - B_{\phi t}  = - \frac{{s^2 a\alpha \left( {\Upsilon + G_{\left( x \right)} y^2  - G_{\left( y \right)} x^2 } \right)}}{{{\cal F}\Xi ^2  + s^2 \left( {{\cal F}\Xi ^2  + \Upsilon } \right)}}\,.
\ee
In the equations above, $\Upsilon  = G_{\left( y \right)}  + G_{\left( x \right)} a^2 \alpha ^2 y^4 $. Again, to obtain the Einstein frame version of (\ref{CmetricXYhet}), one can simply use the relation (\ref{stringTOeinsteinFRAME}). 

Now let us discuss some conditions to guarantee the regularity and correct spacetime signature in the new solutions. However, without performing a delicate analysis on the quite involved metric (\ref{CmetricXYhet}), the Lorentzian nature of seed metrics (\ref{rotcmetric}) and (\ref{CmetricXYkerr}) \cite{Hong:2004dm,Griffiths:2006tk,Griffiths:2005se,Griffiths:2009dfa} guarantees the correct spacetime signature of the solutions (\ref{CKSstringframe}) and (\ref{CmetricXYhet}) respectively. This can be understood from the general form of tetrad component in the transformed solution $e_\mu ^{\left( a \right)} $, which yields $g_{tt} = {\tilde g}_{tt}\Lambda^{-2}$ in the string frame\footnote{Or $g_{tt} = {\tilde g}_{tt}\Lambda^{-1}$ in Einstein frame.}. Furthermore, to have a real and regular dilaton field $\Phi$, we must have $\Lambda >0$, which consequently yields
\be
0 > \Upsilon >  - \frac{{\Xi ^2 {\cal F}}}{{s^2 }}\left( {1 + s^2 } \right)\,.
\ee 

\section{Some aspects of the solution}\label{sec.aspects}

\subsection{Conical singularities}

The seed solution (\ref{rotcmetric}) contains conical singularities on the two axis
\be 
\mathop {\lim }\limits_{x \to  \pm 1} \frac{{2\pi }}{{1 - x^2 }}\sqrt {\frac{{{\tilde g}_{\phi \phi } }}{{{\tilde g}_{xx} }}}  = 1 \pm 2\alpha m + \alpha ^2 a^2 \,.
\ee 
Without introducing any external fields, these singularities cannot be removed at once \cite{Astorino:2016ybm}. To remove one of the nodal singularities, one can perform a scaling to the coordinate\footnote{A coordinate scaling to remove the conical singularity is also performed in the case of magnetized spacetime \cite{Siahaan:2015xia,Siahaan:2016zjw}, where the conical problem comes from the external magnetic field. } $\phi$, which then allows
\be \label{nodal.kerr}
\mathop {\lim }\limits_{x \to 1} \frac{{2\pi }}{{1 - x^2 }}\sqrt {\frac{{{\tilde g}_{\phi \phi } }}{{{\tilde g}_{xx} }}}  = 1\,,~~~{\rm or}~~~\mathop {\lim }\limits_{x \to  - 1} \frac{{2\pi }}{{1 - x^2 }}\sqrt {\frac{{{\tilde g}_{\phi \phi } }}{{{\tilde g}_{xx} }}}  = 1\,,
\ee  
but not altogether simultaneously. The scaling itself incorporate the constants
\be 
C_ \pm   = 1 \pm 2\alpha m + \alpha ^2 a^2 \,,
\ee
where transforming $\phi \to \phi/C_-$ fixes the nodal deficit on $x=-1$, and $\phi \to \phi/C_+$ cures the nodal excess on $x=+1$. The nodal deficit on $x=-1$ is interpreted as a cosmic strut pushing the pair of black holes away each other, while the nodal excess on $x=1$ represents a semi-infinite cosmic string pulling the black hole to infinity. 

Since the seed solution (\ref{rotcmetric}) or (\ref{CmetricXYkerr}) have not been fixed in order to get rid one of the nodal singularities, one can expect that the metric solution resulting from the Hassan-Sen transformation to these seed metrics would suffer the same conical problem. A straightforward check to the solution (\ref{CKSEinsteinframe})
\be \label{nodal.het}
\mathop {\lim }\limits_{x \to  \pm 1} \frac{{2\pi }}{{1 - x^2 }}\sqrt {\frac{{g_{\phi \phi } }}{{g_{xx} }}}  = 1 \pm 2\alpha m + \alpha ^2 a^2 \,.
\ee
The result (\ref{nodal.het}) is independent of the frame being used, i.e. either string frame or Einstein one. Moreover, the outcome (\ref{nodal.het}) is exactly equal to that of accelerating Kerr (\ref{nodal.kerr}), which can be understood from a further little check where
\be 
\mathop {\lim }\limits_{x \to  \pm 1} \frac{{g_{\phi \phi } }}{{{\tilde g}_{\phi \phi } }} = 1\,.
\ee 
Therefore, to get rid of the deficit angle in (\ref{CKSEinsteinframe}), we can employ the same trick as that in accelerating Kerr, i.e. rescaling the $\phi$ coordinate. We prefer to perform $\phi \to \phi/C_-$, hence the metric (\ref{CKSEinsteinframe}) becomes
\[
ds_{E}^2  =  - \frac{{{\cal Q} - {\cal P}a^2 }}{{\Omega ^2 \rho ^2 \Lambda }}\left( {dt - \frac{{a\left( {1 + s^2 } \right)\left( {{\cal Q}{\left(1-x^2\right)}  - \left(r^2+a^2\right) {\cal P}} \right)}}{C_-\left({{\cal Q} - {\cal P}a^2 }\right)}d\phi } \right)^2 
\]
\be \label{CKSEinsteinframeFREEnodalDEF}
+ \frac{{\rho ^2 \Lambda}}{{\Omega ^2 }}\left( {\frac{{dr^2 }}{\cal Q} + \frac{{dx^2 }}{{\cal P}} + \frac{{{\cal QP}}}{{C_-^2 \left({{\cal Q} - {\cal P}a^2 }\right)}}d\phi ^2 } \right)\,.
\ee
Alternatively, we can employ the Hassan-Sen transformation to the accelerating Kerr black hole metric with no conical deficit, and we will arrive at the same result in (\ref{CKSEinsteinframeFREEnodalDEF}). Surely, this scaling to cure one of the conical singularity affects the vector $A_\mu$ and tensor field $B_{\mu\nu}$ in the solution, either by the change of ${\tilde g}_{t\phi}$ and ${\tilde g}_{\phi\phi}$ in the seed solution, or due to the coordinate transformation affecting the component of tensor fields. 

\subsection{Area and temperature}

Now let us study some of black hole's properties based on solutions presented in section \ref{s.1}. Using the Einstein-frame line element (\ref{CKSEinsteinframeFREEnodalDEF}), the area of black hole can be found as
\be 
A_{BH} = \int\limits_0^{2\pi } {d\phi } \int\limits_{ - 1}^1 {dx\left. {\sqrt {g_{xx} g_{\phi \phi } } } \right|_{r = r_ +  } }  = 2\pi \frac{{\left( {1 + s^2 } \right)\left( {r_ + ^2  + a^2 } \right)}}{C_-\left({1 - \alpha ^2 r_ + ^2 }\right)}\,.
\ee 
Setting $s=0$, this area reduces to the area of accelerating Kerr black hole \cite{Appels:2016uha}. Moreover, rewriting this area is term of Kerr-Sen mass \cite{Sen:1992ua,Peng:2016wzr}
\be 
M_{KS}  = \left( {1 + s^2 } \right)m\,,
\ee 
we have
\be 
A_{BH} = \frac{{8\pi M_{KS}~r_ +  }}{C_-\left({1 - \alpha ^2 r_ + ^2}\right)} \,,
\ee 
which is just the area of Kerr-Sen black holes when $\alpha =0$. Since the solution in section \ref{s.1} is interpreted as the accelerating version of Kerr-Sen black hole, then one may expect the black hole contained in the spacetime (\ref{CKSEinsteinframeFREEnodalDEF}) is also rotating just like the generic Kerr-Sen, it can be found that the corresponding angular velocity at the horizon for this black hole is
\be 
\Omega _{BH}  =  \frac{{aC_ -  }}{{\left( {r_ + ^2  + a^2 } \right)\left( {1 + s^2 } \right)}}\,.
\ee 
Interestingly, the acceleration parameter does not appear in this angular velocity, as it should be, just like in the accelerating Kerr \cite{Griffiths:2009dfa}. Using the Killing vector $\zeta ^\mu   = \left[ {1,0,0,\Omega _{BH} } \right]$ which generates the black hole event horizon, one can compute the Coulomb potential of the black hole horizon, i.e.
\be 
\Phi _C  =  - \left. {\zeta ^\mu  A_\mu  } \right|_{r = r_ +  }  =  - \frac{2s}{\sqrt{1+s^2}}\,.
\ee 

To get the temperature, we can make use of the tunneling method \cite{Parikh:1999mf,Banerjee:2008cf,Siahaan:2009qv}. To make the calculation simpler, we consider the tunneling takes place at $x = 1$ axis, even though the obtained result is general \cite{Banerjee:2008cf}. This consideration allows the writing of line element (\ref{CKSstringframe}) to be diagonal, hence for the radial null geodesic $dx = d\phi =0$ we have
\be 
ds^2  =  - f\left( r \right)dt^2  + g^{ - 1} \left( r \right)dr^2 \,,
\ee
where
\[
f\left( r \right) = \frac{\cal Q}{{\Omega ^2 \rho ^2 \Lambda ^2 }} ~,~ g\left( r \right) = \frac{{{\cal Q}\Omega ^2 }}{{\rho ^2 }}\,.
\]
Then the Hawking temperature is given by \cite{Banerjee:2008cf,Siahaan:2009qv}
\be 
T_H  = \left. {\frac{{\sqrt {\left( {\partial _r f} \right)\left( {\partial _r g} \right)} }}{{4\pi }}} \right|_{r = r_ +  } \,.
\ee 
Straightforward calculation yields
\be \label{HawkingTempAccKerrSen}
T_H  = \frac{{\Omega _ + ^2 }}{{\pi \rho _ + ^4 d^3 }}\prod\limits_{j = 1}^3 {c_j } 
\ee 
where
\[
c_1  = \alpha ^2 r_ +  \left( {2r_ + ^4  - 3mr_ + ^3  + 4a^2 r_ + ^2  - 5ma^2 r_ +   + 2a^4 } \right) - m\left( {r_ + ^2  - a^2 } \right)\,,
\]
\[
c_2  = \alpha ^2 mr_ + ^2 \left( {r_ + ^2  - a^2 } \right) - m\left( {r_ + ^2  - a^2 } \right)\,,
\]
\[
c_3  = \alpha r_ +  \left( {a^2  + r_ + ^2  + 2ms^2 r_ +  } \right) + 2m\left( {1 + s^2 } \right)r_ +  \,,
\]
and
\[
d = 2m\left( {1 + s^2 } \right)\left\{ {\alpha \left( {m\left( {r_ +   - r_ -  } \right) + 2m^2  - a^2 } \right) + r_ +  } \right\}\,.
\]
Taking the non-accelerating limit of this temperature, i.e. $\alpha =0$, we get
\[
T_H  = \frac{{r_ +   - m}}{{4\pi mr_ +  \left( {1 + s^2 } \right)}}
\]
which is the Hawking temperature for Kerr-Sen black hole \cite{Ghezelbash:2009gf}. Furthermore, one can observe that the Hawking temperature (\ref{HawkingTempAccKerrSen}) is independent of any frame being considered, either Einstein or string ones. 

Some other main properties of the black hole, that may be extracted from the available solution presented in the previous section are the mass, electric charge, and angular momentum. These are sometime referred as the conserved quantities in the theory, i.e. related to the symmetries possessed by the system. Nevertheless, the traditional approaches in computing these quantities for the non-accelerating Kerr-Sen black holes \cite{Sen:1992ua,Peng:2016wzr,Ghezelbash:2009gf} cannot be used to compute the mass, electric charge, and angular momentum of the accelerating version. The obvious reason is the non-flat nor non-AdS behavior of its asymptotic. In fact, even in Einstein-Maxwell theory, special treatments are needed to get the conserved quantities of an accelerating Kerr-Newman black holes \cite{Astorino:2016xiy,Astorino:2016ybm,Appels:2016uha,Appels:2017xoe}. In appendix \ref{app.charges}, we highlight an attempt to get the conserved charges for spacetime solution in the low energy heterotic string theory.

\section{Discussions}\label{s.7}

In this paper we have obtained new solutions in the low energy of heterotic string theory, describing a pair of accelerating, rotating, and charged black holes in this theory. In principle, we can get the accelerating Kerr-Sen-NUT solution, which is analogous to the Kerr-Newman-NUT spacetime in Einstein-Maxwell theory. Nevertheless, to avoid the pathological conic singularity due to the presence of NUT parameter, we focus only in the case without this NUT charge. Nevertheless, for interested readers, solutions incorporating the NUT parameter are presented in app. \ref{app.NUT}. In obtaining these solutions, we have employed the Hassan-Sen transformation which can be viewed as a generating solutions method in the low energy limit of heterotic string theory. We also manage to write down a general form of fields solutions in low energy heterotic string theory as some functions of the axially symmetric and stationary vacuum Einstein spacetime solutions. Some properties of the black hole solution such as area, angular velocity, and Hawking temperature are given. However, works presented in this paper are limited to the seed metric which solves the vacuum Einstein equations.

Nonetheless, due to the ``strange'' asymptotics of the obtained solution and more involved equation of motions for the fields, we have not been able to present the conserved charges in the system. To work this out, we may require the integrability consideration in the covariant phase approach, like the one considered in the Einstein-Maxwell theory \cite{Astorino:2016xiy}. This line of research is under investigation, where the covariant phase method to get the conserved charges related to Kerr-Sen spacetime has been discussed in \cite{Ghezelbash:2009gf}. In this paper, we also present a new solution describing the non-accelerating Kerr-Sen-NUT spacetime, whose further physical properties investigations are of particular interests of the reader. Finally, in the line of holographic researches \cite{Ghezelbash:2009gf,Astorino:2016xiy,Anabalon:2018ydc}, holography studies related to the new accelerating black hole solutions reported in this work is also a promising future project.

 \section*{Acknowledgement}
 
{I thank Zoltan Bajnok for his supports, and Chao Wu for the discussions. I also thank my colleagues from Physics Department of Parahyangan Catholic University. This work is supported by Tempus Public Foundation.}

\appendix

\section{Solution with NUT parameters}\label{app.NUT}

In the main body of this paper, we avoid discussing the solution with NUT charge since its pathological conic singularity which comes from the NUT parameter in the solution. Nevertheless, since the accelerating Kerr-NUT metric \cite{Griffiths:2009dfa} solves the vacuum Einstein equation, in principle one can also employ the Hassan-Sen transformation to this solution in generating the set of field solutions representing the accelerating Kerr-Sen-NUT system. Employing the general form of fields (\ref{tetrad.HassanSen}) - (\ref{secondrank.HassanSen}) to the seed spacetime solution (\ref{metricrotCNUT}) yields
\[
ds_{string}^2  =  - \Xi \left( {dt - \frac{{Qa\left( {1 - x} \right)\left( {a\left( {1 + x} \right) + 2l} \right) - P\omega ^2 \left( {r^2  + \left( {l + a} \right)^2 } \right)}}{{a\left( {Q - P\omega ^2 } \right)}}d\phi } \right)^2 
\]
\be \label{metric.acc.KerrSenNUT}
+ \frac{{\varrho ^2 }}{{\Omega ^2 }}\left( {\frac{{dr^2 }}{Q} + \frac{{a^2 dx^2 }}{{\omega ^2 P}}} \right) + \frac{{QP\omega ^2 \varrho ^2 }}{{\Omega ^2 a^2 \left( {Q - P\omega ^2 } \right)}}d\phi ^2 \,,
\ee
where
\be 
\Xi = \frac{{\left( {Q - P\omega ^2 } \right)\Omega ^2 \varrho ^2 }}{{\left( {\Omega ^2 \varrho ^2 \left( {1 + s^2 } \right) - s^2 \left( {Q - P\omega ^2 } \right)} \right)^2 }}\,.
\ee 
The non-gravitational fields accompanying the above metric are
\[
A_\mu  dx^\mu   = \frac{{2s\sqrt {1 + s^2 } }}{{\Omega ^2 \varrho ^2 \left( {1 + s^2 } \right) - s^2 \left( {Q - P\omega ^2 } \right)}}\left[ {\left( {\Omega ^2 \varrho ^2  - Q + P\omega ^2 } \right)dt} \right.
\]
\be 
\left. { + \left( {Q\left( {1 - x} \right)\left( {a\left( {1 + x} \right) + 2l} \right) - \frac{{P\omega ^2 }}{a}\left( {r^2  + \left( {l + a} \right)^2 } \right)} \right)d\phi } \right]\,,
\ee
\be 
\Phi  =  - \ln \left[ {\frac{{\Omega ^2 \varrho ^2 \left( {1 + s^2 } \right) - s^2 \left( {Q - P\omega ^2 } \right)}}{{\Omega ^2 \varrho ^2 }}} \right]\,,
\ee
and
\be \label{Btp.accKerrSenNUT}
B_{t\phi }  =  - B_{\phi t}  = \frac{{Qa\left( {1 - x} \right)\left( {a\left( {1 + x} \right) + 2l} \right) - P\omega ^2 \left( {r^2  + \left( {l + a} \right)^2 } \right)}}{{a\left( {\Omega ^2 \varrho ^2 \left( {1 + s^2 } \right) - s^2 \left( {Q - P\omega ^2 } \right)} \right)}}\,.
\ee
Setting $\alpha = 0$, $k = \left(a^2-l^2\right)\omega^{-2}$, $\epsilon = 1$, $\omega = a$, and $n=l$ in this solution leads to the Kerr-Sen-NUT system. Explicitly, to write down the Kerr-Sen-NUT solution, one simply replace
\be 
\Omega  = 1~~,~~P = 1 - x^2 ~~,~~Q = r^2  - 2mr + a^2  - l^2 \,,
\ee 
in the solutions (\ref{metric.acc.KerrSenNUT}) - (\ref{Btp.accKerrSenNUT}). This Kerr-Sen-NUT system resembles the well known Kerr-Newman-NUT solution in Einstein-Maxwell theory \cite{Griffiths:2009dfa}.

\section{Conserved charges}\label{app.charges}

In \cite{Ghezelbash:2009gf}, the author managed to compute the central charge associated to the conformal symmetry of the near region extremal Kerr-Sen black holes. This was done to established the Kerr/CFT correspondence in the low energy limit of heterotic string theory. The conserved charge employed in this work is that from the covariant phase framework, or sometime referred as the Barnich-Brandt method \cite{Barnich:2001jy}. The charge reads
\be \label{charge.sym}
Q_{\left( {\xi ,\lambda ,\psi } \right)}  = \frac{1}{{8\pi }}\int {dS_{\mu \nu } \left( {k_{\left( g \right)}^{\mu \nu }  + k_{\left( \Phi  \right)}^{\mu \nu }  + k_{\left( A \right)}^{\mu \nu }  + k_{\left( B \right)}^{\mu \nu } } \right)} \,,
\ee 
where
\be \label{kg}
k_{\left( g \right)}^{\mu \nu }  = \xi ^\mu  \nabla _\sigma  h^{\nu \sigma }  - \xi ^\mu  \nabla ^\nu  h - \xi _\sigma  \nabla ^\mu  h^{\nu \sigma }  - \frac{h}{2}\nabla ^\mu  \xi ^\nu   + h^{\mu \sigma } \nabla _\sigma  \xi ^\nu   - \frac{1}{2}h^{\sigma \mu } \left( {\nabla ^\nu  \xi _\sigma   + \nabla _\sigma  \xi ^\mu  } \right)\,,
\ee
\[
k_{\left( A \right)}^{\mu \nu }  = \frac{1}{{2\pi }}\left[ {\left( { - \frac{h}{2}F^{\mu \nu }  + 2F^{\mu \sigma } h_\sigma ^\nu   - f^{\mu \nu } } \right)\left( {\xi ^\sigma  A_\sigma   + \lambda } \right) - F^{\mu \nu } \xi ^\sigma  a_\sigma   - 2F^{\sigma \mu } \xi ^\nu  a_\sigma  } \right]
\] 
\be\label{kA} ~~~~- \frac{1}{2}a^\mu  \left( {\xi _\sigma  F^{\nu \sigma }  + \partial ^\nu  \left( {\xi ^\sigma  A_\sigma   + \lambda } \right)} \right)\,,\ee
\be \label{kPhi}
k_{\left( \Phi  \right)}^{\mu \nu }  = \frac{2}{3}\varphi \xi ^\mu  \partial ^\nu  \Phi \,,
\ee 
and
\[ 
k_{\left( B \right)}^{\mu \nu }  = \frac{1}{3}\left( {\delta _\rho ^\kappa  \delta _\beta ^\sigma  b_{\lambda \alpha }  + \delta _\rho ^\kappa  \delta _\alpha ^\sigma  b_{\beta \lambda }  + \delta _\rho ^\kappa  \delta _\lambda ^\sigma  b_{\alpha \beta } } \right)\xi ^\lambda  H^{\mu \nu \rho }  
\]
\[+ \frac{2}{3}\left[ {\frac{1}{2}\xi ^\lambda  b_{\rho \lambda } H^{\mu \rho \nu }  + \left( {\frac{1}{2}\xi ^\lambda  B_{\rho \lambda }  + {\cal B} _\rho  } \right)\left( {{\mathfrak{h}}^{\mu \rho \nu }  + \frac{h}{2}H^{\mu \rho \nu } } \right)} \right]\]
\be \label{kB}
+ \frac{1}{2}b^{\alpha \mu } \left( {B_\rho ^\nu  \partial _\alpha  \xi ^\rho   + B_{\rho \alpha } \partial ^\nu  \xi ^\rho   + \xi ^\rho  \partial _\rho  B_\alpha ^\nu   + \partial ^\nu  {\cal B} _\alpha   - \partial _\alpha  {\cal B} ^\nu  } \right)\,.
\ee
In equation above, integration over the surface $S$ is defined as
\be 
dS_{\mu \nu }  = \frac{{\sqrt { - g} }}{4}\varepsilon _{\mu \nu \alpha \beta } dx^\alpha   \wedge dx^\beta  \,,
\ee
with $\varepsilon_{0123}=1$. Furthermore, 
\be 
h_{\mu \nu \rho }  = \partial _\mu  b_{\nu \rho }  + \partial _\rho  b_{\mu \nu }  + \partial _\nu  b_{\rho \mu }  - \frac{1}{4}\left( {a_\mu  f_{\nu \rho }  + a_\rho  f_{\mu \nu }  + a_\nu  f_{\rho \mu } } \right)\,,
\ee
and 
\be 
f_{\mu \nu }  = \partial _\mu  a_\nu   - \partial _\nu  a_\mu  \,.
\ee

In (\ref{kA}) and (\ref{kB}), the functions $\lambda$ and ${\cal B}_\alpha$ are related to the gauge freedom of $A_\mu$ and $B_{\mu\nu}$, 
\be
A_\mu   \to A_\mu   + \partial _\mu  {\lambda} \,,
\ee
\be
B_{\mu \nu }  \to B_{\mu \nu }  + \partial _\mu  {\cal B}_\nu   - \partial _\nu  {\cal B}_\mu  \,.
\ee

The fields denoted by lower case letters, i.e. $h_{\mu\nu}$, $a_{\mu}$, $b_{\mu\nu}$ are variation of the spacetime metric $g_{\mu\nu}$, vector field $A_{\mu}$, and second rank antisymmetric field $B_{\mu\nu}$ respectively, depending on what quantities change in the theory.  For example, and this is the consideration in this paper, we let the ``Kerr mass'' can vary as $m\to m+\delta m$, thence 
\be 
h_{\mu \nu }  = \frac{{\partial g_{\mu \nu } }}{{\partial m}}\delta m~~,~~a_\mu   = \frac{{\partial A_\mu  }}{{\partial m}}\delta m~~,~~b_{\mu \nu }  = \frac{{\partial B_{\mu \nu } }}{{\partial m}}\delta m~~,~~\varphi  = \frac{{\partial \Phi }}{{\partial m}}\delta m\,,
\ee
where $\varphi$ is related to the dilaton. Raising and lowering index in the expressions above are performed by using $g_{\mu\nu}$, for example $h = g_{\mu\nu}h^{\mu\nu}$. 

In the case of non-accelerating Kerr-Sen black holes, the charge (\ref{charge.sym}) gives the mass and angular momentum after considering the Killing vectors $\xi _{\left( t \right)}^\mu   = \left[ { - 1,0,0,0} \right]$ and $\xi _{\left( \phi  \right)}^\mu   = \left[ {0,0,0,1} \right]$ respectively. Moreover, the author of \cite{Peng:2016wzr} managed to obtain the mass and angular momentum of Kerr-Sen black hole by using the generalized ADT method. This can be understood since the covariant phase and ADT methods are basically equivalent \cite{Hyun:2014sha}. However, we find that both methods fail to give the corresponding mass, electric charge, and angular momentum for accelerating Kerr-Sen black hole, where the outcomes are divergent.

\end{document}